\documentclass[journal]{IEEEtran}
\usepackage{siunitx} %
\usepackage{multirow}
\usepackage{xcolor}
\definecolor{redcolor}{rgb}{1, 0, 0}
\usepackage{graphicx}
\usepackage{psfrag}  
\usepackage{subfigure}
\usepackage{tabularx}
\usepackage{longtable}
\usepackage{threeparttable}
\usepackage{textcomp}
\usepackage{amsmath, bm}

\IEEEoverridecommandlockouts
\begin{document}
\bibliographystyle{IEEEtran}
\bstctlcite{IEEEexample:BSTcontrol}
\nocite{*}
\title{Analysis and Design Considerations for Achieving the Fundamental Limits of Phase Noise in mmWave Oscillators with On-Chip MEMS Resonator}
\author{Abhishek~Srivastava,~\IEEEmembership{Member,~IEEE,}
        Baibhab~Chatterjee,~\IEEEmembership{Student Member,~IEEE,}
        Udit~Rawat,~\IEEEmembership{Student Member,~IEEE,}
        Yanbo~He,~\IEEEmembership{Student Member,~IEEE,}
        Dana~Weinstein,~\IEEEmembership{Senior Member,~IEEE,}
        and~Shreyas~Sen,~\IEEEmembership{Senior Member,~IEEE,}
        \thanks{Authors are associated with the Department of Electrical and Computer Engineering, Purdue University, IN, USA.} %
        }%
\maketitle
\begin{abstract}
Very small electromechanical coupling coefficient 
in micro-electromechanical systems (MEMS) or acoustic resonators is quite of a concern for oscillator performance, specially at mmWave frequencies. 
This small coefficient is the manifestation of the small ratio of motional capacitance to static capacitance in the resonators. 
This work provides a general solution to overcome the problem of relatively high static capacitance at mmWave frequencies and presents analysis and design techniques for achieving 
extremely low phase noise and a very high figure-of-merit (FoM) in an on-chip MEMS resonator based mmWave oscillator. 
The proposed analysis and techniques are validated with design and simulation of a 30 GHz oscillator with MEMS resonator having quality factor of 10,000 in 14 nm GF technology. Post layout simulation results show that it achieves a phase noise of -132 dBc/Hz and FoM of 217 dBc/Hz at offset of 1 MHz. 
\end{abstract}
\renewcommand\IEEEkeywordsname{Keywords}
\IEEEpeerreviewmaketitle
\section{Introduction}
Increasing demands of 
smaller foot-print for multi-channel integration 
and reduced bill-of-material (BoM) in %
wireless communication technologies, such as 5G, 
seek
low power radio frequency (RF) transceivers with 
monolithic resonators as compared to the conventional bulky off-chip crystals.
For multi-channel integration, where large arrays of transceivers are used on the same chip, local carrier generation in each transceiver is desirable %
than distributing a common clock over entire chip in order to 
reduce the clock routing power 
and have better control over the phases of the RF carriers used for channel integration.  
Power can be further saved if RF carriers are generated without using a phase lock loop (PLL) 
in each transceiver.  
However, 
problem with PLL-less RF synthesis is the compromised phase noise, which requires alternate low phase noise techniques.  %
With the advancements in the MEMS technologies and recently reported high-frequency, high-Q  ($>1,000$) resonators 
\cite{Fischer_2015}-\cite{rft_isscc2018}, 
it seems possible to build low phase noise 
PLL-free 
oscillators with on-chip MEMS resonators for direct RF generation 
even at mmWave frequencies. 

\begin{figure}%
\centering
\includegraphics[scale=0.43]{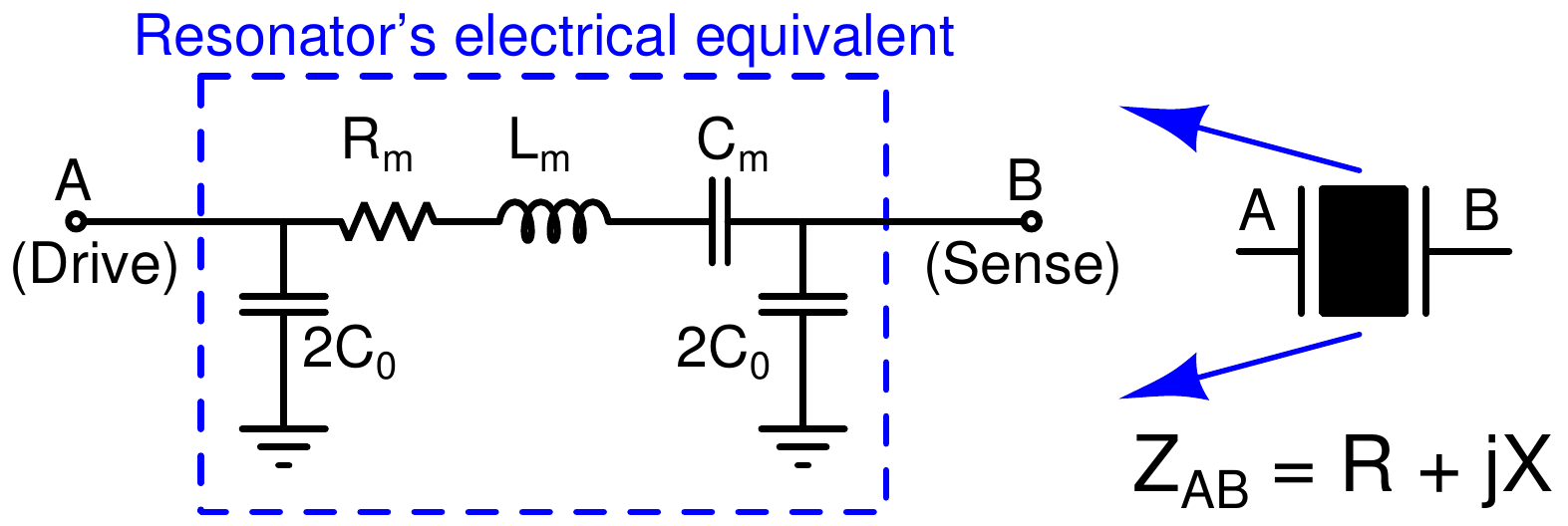}
\vspace{-4mm}
\caption{Butterworth Van Dyke model of MEMS resonator}
\label{fig_reso_bvd}
\vspace{-3mm}
\end{figure}

Fig. \ref{fig_reso_bvd} shows the Butterworth–Van Dyke model of a MEMS resonator, where $R_m$, $L_m$ and $C_m$ are termed as motional resistance, inductance and capacitance, respectively. 
$C_0$ is the static capacitance due to the device structure and geometry at the driving and sensing ports. 
Table \ref{tab_params} lists parameters of few resonators. 
\begin{table}
\vspace{-2mm}
\begin{center}
\footnotesize
  \footnotesize{\caption{Electrical Parameters of Different Resonators}}
  \vspace{-6mm}
  \label{tab_params}
    \resizebox{1\linewidth}{!}{
    \begin{tabular}{|>{\raggedright}p{1.1cm}|>{\raggedright}p{.6cm}|>{\raggedright}p{.8cm}|>{\raggedright}p{1.15cm}|>{\raggedright}p{1cm}|>{\raggedright}p{.9cm}|>{\raggedright}p{.8cm}|>{\raggedright}p{.8cm}|}
        \hline
         \textbf{Frequency} &\textbf{ $\bm Q$} &\textbf{$\bm R_m$} & \textbf{$\bm L_m$} & \textbf{$\bm C_m$} & \textbf{$\bm C_0$}  & \textbf{$\bm {|X_{C_0}|}$} & $\bm {\frac{|X_{C_0}|}{R_m}}$\tabularnewline
         \hline
         45 MHz &$10^5$ & 12.3 $\Omega$ & 4.4 mH& 2.895 fF& 4 pF& 884 $\Omega$ & $\approx 72$ \tabularnewline
(Quartz)\cite{crystal_reso} & & & & & & &\tabularnewline
         \hline
         400 MHz & 16,400 & 14 $\Omega$ & 97.5 $\mu$H & 1.594 fF& 2.1 pF& 190 $\Omega$ & $\approx 13$\tabularnewline
         (SAW)\cite{saw_reso} & & & & & & &\tabularnewline
         \hline
         2.4 GHz & 1600& 1.04 $\Omega$ & 107.2 nH& 38.99 fF & 1.29 pF& 51.3 $\Omega$& $\approx 50$ \tabularnewline
         (FBAR)\cite{fbar_2013} & & & & & & &\tabularnewline
         \hline
         30 GHz & 10,000& 332 $\Omega$ & 17.59 $\mu$H & 1.6 aF & 16 fF& 331 $\Omega$ & $\approx 1$ \tabularnewline
         (RFT) \dag & & & & & & &\tabularnewline
         \hline                           
        \end{tabular}
        }
\begin{tablenotes}
 \footnotesize
\raggedright
\vspace{-1mm}
\item \dag This work
\vspace{-2mm}
 \end{tablenotes}
\end{center}
\vspace{-7mm}
\end{table}
Usually MEMS resonators have very low electromechanical coupling coefficient ${k_t}^2$ ($=C_m/C_0$), which 
signifies a low efficiency of energy transfer between electrical and mechanical domains. 
At mmWave frequencies, $C_0$ exacerbates the problem of low ${k_t}^2$, which can lead to higher phase noise and lower figure-of-merit (FoM) of the oscillator, if not addressed appropriately. 

In this work, 
towards the goal of building low phase noise mmWave oscillator with on-chip high-Q MEMS resonator, 
we present: 
1) design challenges due to $C_0$ and their general solutions,
2) theoretical analysis for fundamental limits of phase noise and FoM and 
3) general design technique to achieve high FoM. %
To validate the proposed analysis and design approach, a 30 GHz oscillator %
is designed and simulated in 14-nm Global Foundry (GF) process. 
While the important insights presented in this work %
are shown with a high-Q Resonant FinFET (RFT) MEMS device \cite{rft_isscc2018}, they 
can be applied in general for building extremely low phase noise mmWave oscillator with any high-Q monolithic MEMS resonator. %

This paper is organized as follows. Section \ref{mems_resonator} presents the design details of 30 GHz MEMS resonator used in this work. Sections \ref{c0_challenge} and \ref{c0_compensation}, discuss the challenges due to $C_0$ and their solutions, respectively. 
Section \ref{pn_model} presents the analysis for fundamental phase noise and FoM limits with general design technique.
Section \ref{impl_result} presents the implementation details with simulation results followed by the conclusion of the paper. 

\section{RFT : mmWave MEMS Resonator}
\label{mems_resonator}
\begin{figure}
\centering
\includegraphics[scale=0.22]{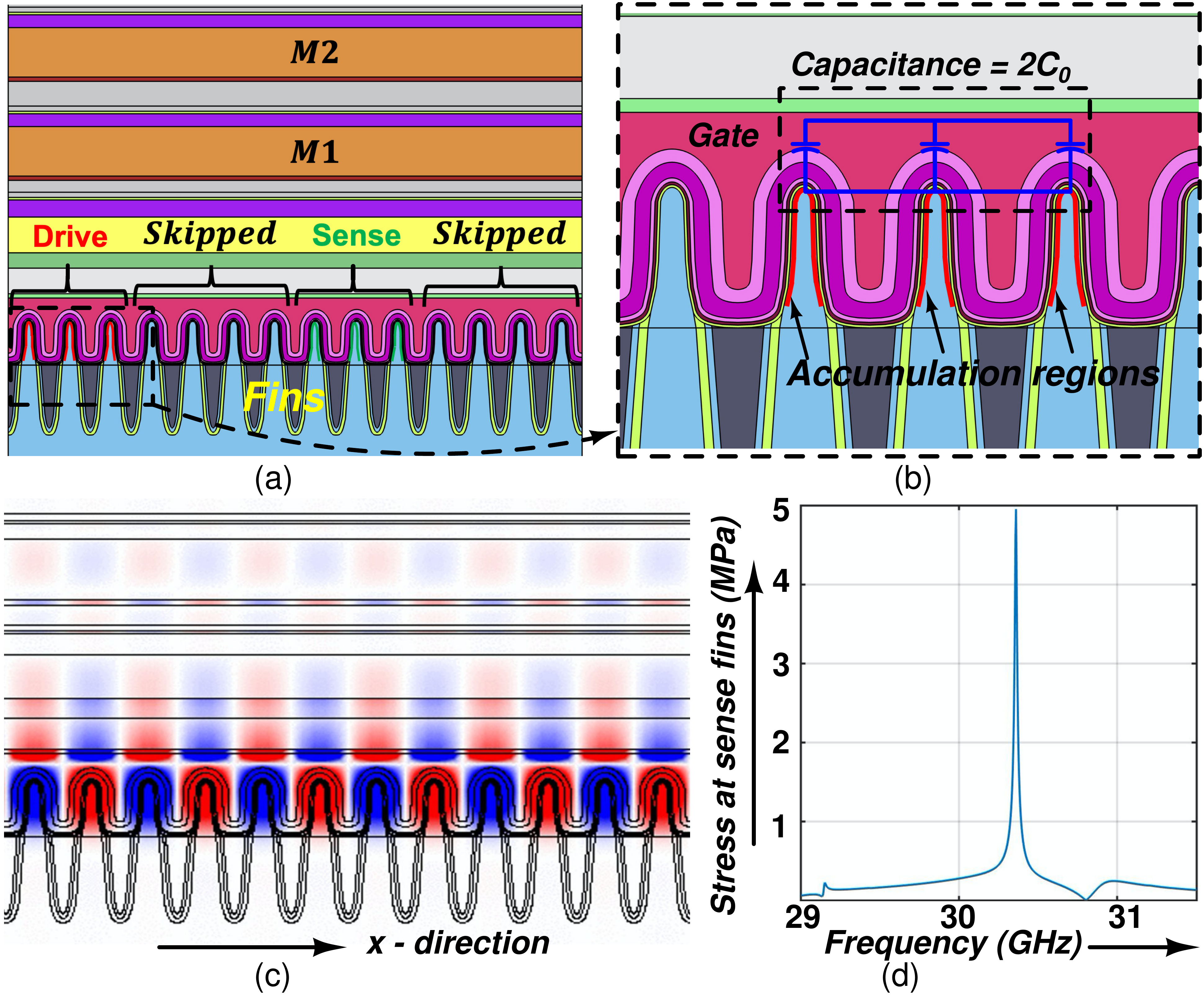}
\caption{(a) Cross-sectional view of a unit cell of the 30 GHz RFT resonator showing the drive and sense transducers along with the interleaved skipped fins for satisfying DRC requirements (b) zoomed-in view of the cross section depicting the physical origin of the capacitance $C_0$ (c) mechanical mode shape showing regions of maximum (red) and minimum (blue) stress in the x-direction (d) frequency domain FEM simulation result depicting the x-stress in the sense transistor fins}
\label{fig_rft_all}
\vspace{-3mm}
\end{figure}
Fig. \ref{fig_rft_all}(a) depicts the cross sectional view of a unit cell of the 30 GHz RFT MEMS resonator. An acoustic waveguide capable of sustaining mechanical modes of vibration is created by repeating this unit cell in the x-direction. In order to confine the waveguide mode of interest in the fin region, back end of Line (BEOL) metal layers are used as Bragg reflectors. 
Each unit cell consists of separate, 3-fin drive and sense MOSCAPs ($=2C_0$) in which the gate and the shorted source-drain form the two plates of the capacitor as shown in Fig. \ref{fig_rft_all}(b). 
The drive MOSCAPs are used as capacitive transducers for actuation, 
which are biased in the accumulation region for maximizing the transduction efficiency. 
Moreover, multiple drive units connected in parallel are needed for efficient electromechanical coupling, which results in a large $C_0$. %

The mode shape of RFT is shown in Fig. \ref{fig_rft_all}(c) depicting the alternating polarity of the mechanical stress in the x-direction at each fin. A Finite Element Method (FEM) analysis based simulation yields the x-stress in the fins of the sense MOSCAPs with the maximum occurring at the resonance frequency of 30.3 GHz as shown in Fig. \ref{fig_rft_all}(d). Capacitive sensing is employed at the sense MOSCAPs which are also biased in the accumulation region. The sense capacitance gets modulated by the mechanical vibration in the resonant cavity and the corresponding signal generated is read out. 

To obtain the maximum strength of the output signal from the sense MOSCAPs, $C_0$ needs to be as large as possible. The resonator performance thus places a limit on the value of $C_0$ which cannot be reduced without severely degrading the transduction efficiency of both the drive as well as sense transducers. $C_0$ is evaluated to be 16 fF through an AC simulation of the layout extracted netlist of the RFT device.
\section{Challenges In Oscillator Design Due To $C_0$ At mmWave Frequencies}
\label{c0_challenge}
\begin{figure}%
\centering
\includegraphics[scale=0.32]{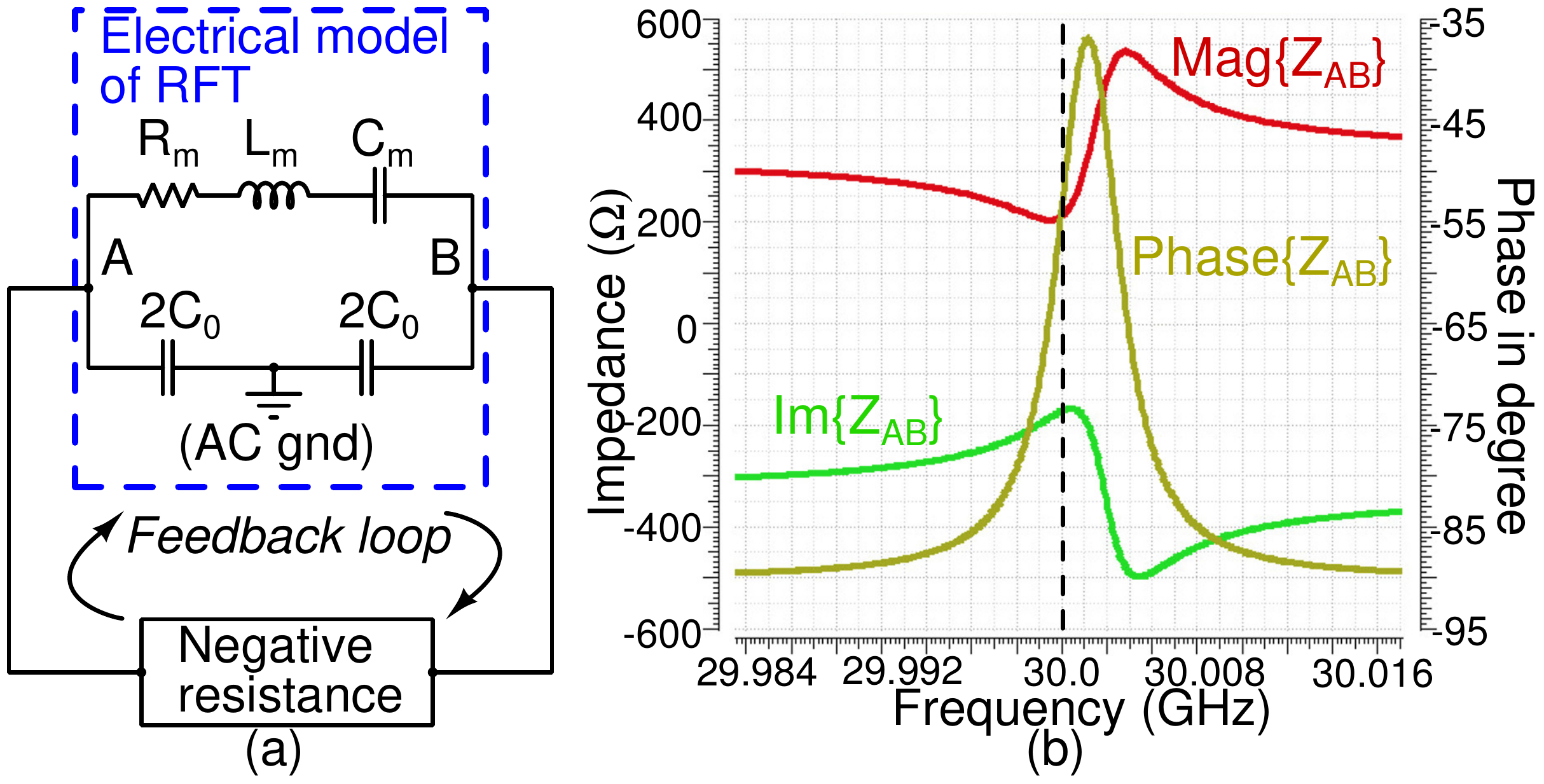}
\caption{(a) Oscillator depiction with RFT, (b) impedance vs frequency plots for RFT's electrical model shown in (a)}
\label{fig_reso_phase}
\end{figure}
Electrical model of the RFT device is shown in Fig. \ref{fig_reso_phase}(a). It exhibits two resonance modes - series and parallel. 
The series ($f_s$) and parallel ($f_p$) resonant frequencies are defined in Eq. (\ref{eq_fs_fp})
 \cite{vitoz_crystal}. 
\begin{equation}
   \label{eq_fs_fp}
   \small
    f_{s} = \frac{1}{2{\pi}\sqrt{L_{m}C_{m}}}; f_{p} = f_{s}\bigg\{1 + \frac{C_{m}}{2C_{0}}\bigg\}
\end{equation} 
Impedance ($Z_{AB}$) across the resonator 
and its phase ($\phi _{Z_{AB}}$) can be given by the equations (\ref{eq_imp_reso}) and (\ref{eq_phase_reso}), respectively. 
\begin{equation}
\label{eq_imp_reso}
\small
	Z_{AB} = \frac{(1-\omega^2L_mC_m) + j\omega R_{m}C_{m}}{-\omega^2R_{m}C_{m}C_{0} - j\{\omega^3L_mC_{m}C_{0} - \omega(C_{m}+C_{0})\}} 
\end{equation}
\begin{multline}
\small
\label{eq_phase_reso}
\phi _{Z_{AB}} = \pi + tan^{-1}\frac{\omega R_m C_m}{(1-\omega^2L_mC_m)}\ - \\ 
tan^{-1}\frac{\omega^3L_mC_{m}C_{0} - \omega(C_{m}+C_{0})}{\omega^2R_{m}C_{m}C_{0}}
\end{multline}
As depicted in Fig. \ref{fig_reso_phase}(a), oscillators with MEMS resonator are conventionally built with a negative resistance in a close loop at $f_s$ or $f_p$, 
where $\phi _{Z_{AB}} = 0$\textdegree\ and resonator behaves as a resistor \cite{vitoz_crystal}. 
However, at mmWave frequencies $C_0$ poses some challenges discussed below, which 
makes it impossible to build oscillators by directly applying the conventional methods. 
\paragraph{Non-zero phase}
From Eq. \ref{eq_phase_reso}, by equating $\phi _{Z_{AB}}$ to 0\textdegree, following condition (\ref{eq_c0}) is obtained.
\begin{equation}
\label{eq_c0}
C_0 = \frac{C_m(\omega^2L_mC_m-1)}{\omega^2{R_{m}}^2{C_m}^2 + (\omega^2L_mC_m-1)^2}
\end{equation}
From (\ref{eq_c0}) and the motional parameters of RFT device shown in table \ref{tab_params}, value of $C_0$ required for 0\textdegree\ phase shift at 30 GHz is about %
1.6 fF. However, this small value of $C_0$ is not feasible as explained earlier in section \ref{mems_resonator}. 
Magnitude and phase of the 30 GHz RFT device are also plotted in Fig. \ref{fig_reso_phase}(b), 
which shows that the
impedance is capacitive near 30 GHz and the resonator does not provide 0\textdegree\ phase shift and therefore, it can not be directly used to build an oscillator. 
\paragraph{Signal loss at mmWave frequencies}
As shown in the table \ref{tab_params}, while static capacitance $C_0$ reduces as frequency increases, $X_{C_0}$ ($= \frac{1}{j\omega C_0}$) also reduces and becomes comparable to $R_m$ for the 30 GHz resonator. 
Therefore at $f_s$, about half of the signal will flow to ground through $C_0$, which will pose higher driving requirements to build the oscillator. 
From the foregoing discussions, some fundamental questions arise for utilizing on-chip MEMS resonators at mmWave frequencies: 
1) how to get 0\textdegree\,phase-shift, 2) how to avoid signal loss and 3) is there any advantage in phase noise and FoM as compared to conventional LC oscillators. These questions are answered in sections \ref{c0_compensation} and \ref{pn_model}. 
\section{$C_0$ Compensation}
\label{c0_compensation}
\begin{figure}
\centering
\includegraphics[scale=0.24]{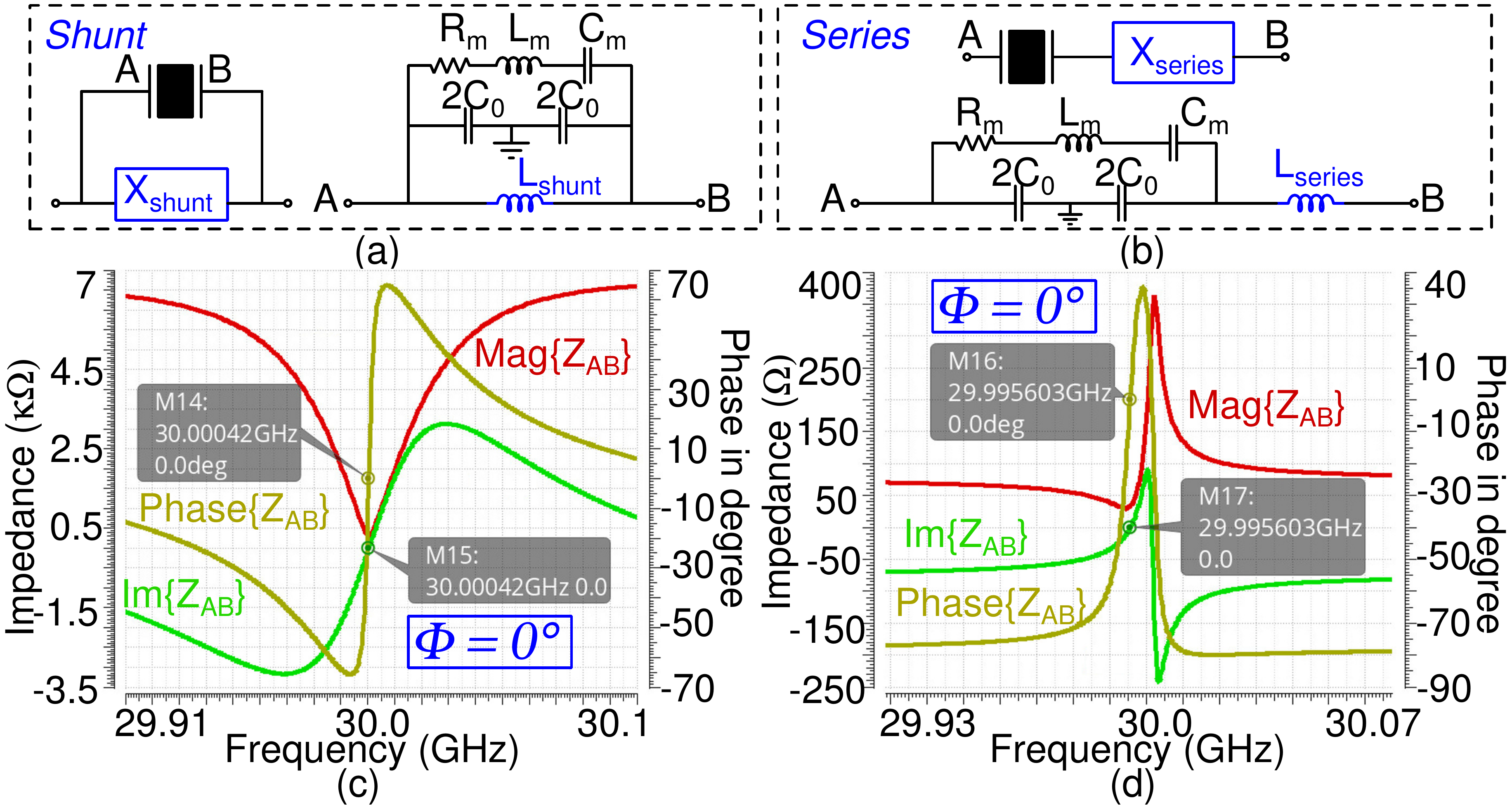}
\caption{(a) Shunt compensation, (b) series compensation, (c) impedance plots for (a), (d) impedance plots for (c)}
\label{fig_c0_cancel}
\end{figure}

In order to make $\phi _{Z_{AB}} = 0$\textdegree\ at the frequency of interest ($f_0$), $Im\{Z_{AB}\}$ should be 0 at $f_0$. 
Since ${Z_{AB}}$ is capacitive near 30 GHz (Fig. \ref{fig_reso_phase}(b)), 
it can be cancelled with an inductor in series or shunt as shown in figures \ref{fig_c0_cancel}(a) and \ref{fig_c0_cancel}(b), respectively. 
Phase and magnitude of the impedances of the two schemes are plotted in figures \ref{fig_c0_cancel}(c) and \ref{fig_c0_cancel}(d), respectively, which show that $\phi _{Z_{AB}} = 0$\textdegree\ near 30 GHz. 
At resonance, 
scheme of Fig. \ref{fig_c0_cancel}(b) solves the problem of non-zero phase, however, it does not solve the problem of signal loss to ground through $C_0$. 
Therefore shunt inductor compensation (Fig. \ref{fig_c0_cancel}(a)) is a better choice to resonate out $C_0$, which solves both the problems.

\begin{figure}%
\centering
\includegraphics[scale=0.167]{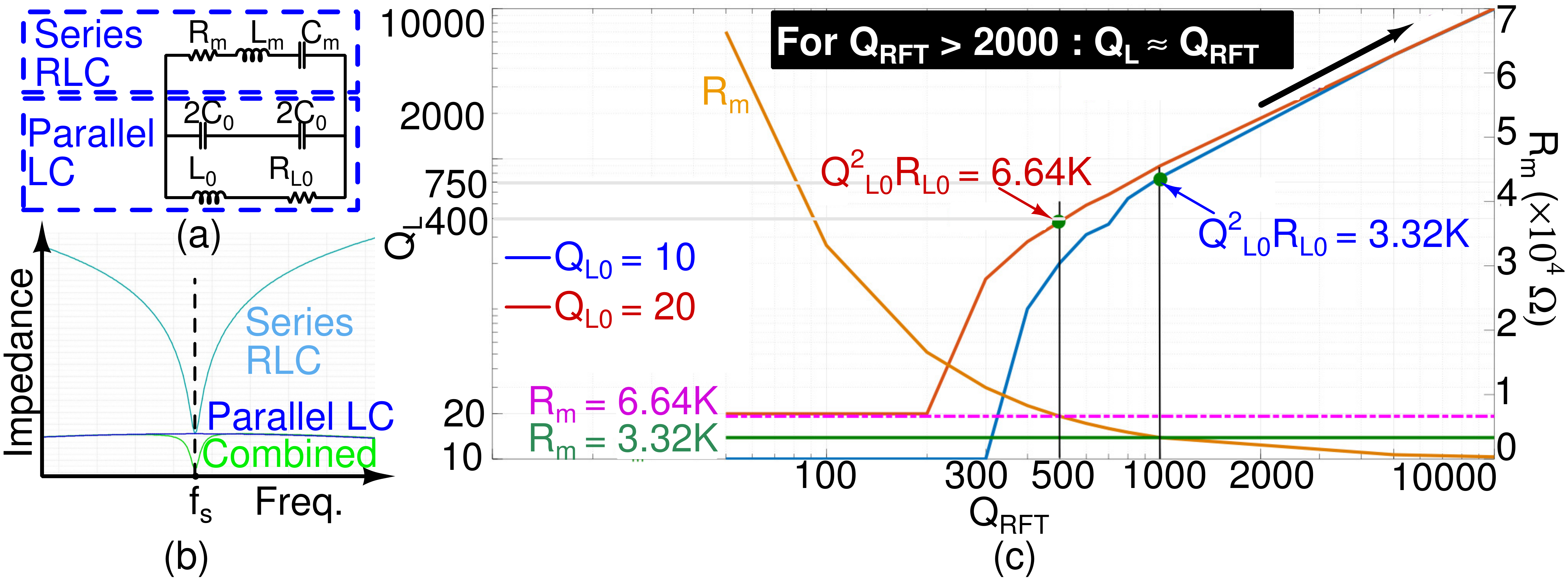}
\caption{(a) Shunt inductor compensated MEMS resonator, (b) impedance plot showing parallel combination of $R_mL_mC_m$ and $L_0C_0$ branches, (c) $Q_{L}$ vs $Q_{RFT}$}
\label{fig_QL_QRESO}
\end{figure}

The shunt inductor ($L_0$) compensated resonator can also be thought of as a parallel combination of series $R_mL_mC_m$ and the parallel $L_0C_0$ branches 
as shown in Fig. \ref{fig_QL_QRESO}(a) with an overall AC response shown in Fig. \ref{fig_QL_QRESO}(b). 
Q-factor of on-chip inductors ($Q_{L_0}$) are usually limited ($<30$), %
therefore Fig. \ref{fig_QL_QRESO}(a) 
exhibits a loaded Q-factor ($Q_L$). 
However, as shown  in Fig. \ref{fig_QL_QRESO}(b), the overall frequency response (green coloured) is governed by the high-Q motional branch near $f_s$.
Fig. \ref{fig_QL_QRESO}(c) shows $Q_L$ as a function of MEMS resonator's Q-factor ($Q_{RFT}$) for two values of $Q_{L_0}$ - 10 and 20.
As shown in the figure, for lower values of $Q_{RFT}\ (<1000)$, there is a significant effect of loading and $Q_L$ is low. However, for  $Q_{RFT} >$ 1000 loading reduces and for $Q_{RFT} >2000$, $Q_L \approx Q_{RFT}$. 
Therefore, as long as $Q_{RFT} >2000$, %
on-chip shunt inductor with inherent low $Q_{L_0}$ ($\leq 20$) will not cause any significant loading and 
extremely low phase noise oscillators can be built.  
Phase noise analysis and design technique for the oscillator are discussed in the following section. 
\section{Phase Noise Analysis And General Design Technique For mmWave Oscillator}
\label{pn_model}
For phase noise analysis, the oscillator with shunt-inductor compensated MEMS resonator can be considered as an LC feedback oscillator with the resonance characteristics as shown in Fig. \ref{fig_QL_QRESO}(b). %
Leeson's proportionality defined in Eq. (\ref{eq_leeson}) can be used to capture the phase noise ($\mathcal{L}\{\Delta f\}$) of the  oscillator at an offset of $\Delta f$ with center frequency $f_0$ \cite{leeson_pn}.  
\begin{equation}
\label{eq_leeson}
\small
\mathcal{L}\{\Delta f\} = F \frac{4kTR_m}{V_{OSC}^2} \bigg(\frac{f_0}{2Q_L \Delta f}\bigg)^2
\end{equation}
In Eq. (\ref{eq_leeson}) , F is the noise factor of oscillator, which is defined as the total oscillator phase noise normalized to phase noise due to the MEMS resonator loss ($R_m$), $k$ is the Boltzmann's constant and T is the temperature in Kelvin. 
\subsection{Intuitive Analysis}
\label{pn_model_intuitive}
\begin{figure}
\centering
\includegraphics[scale=0.37]{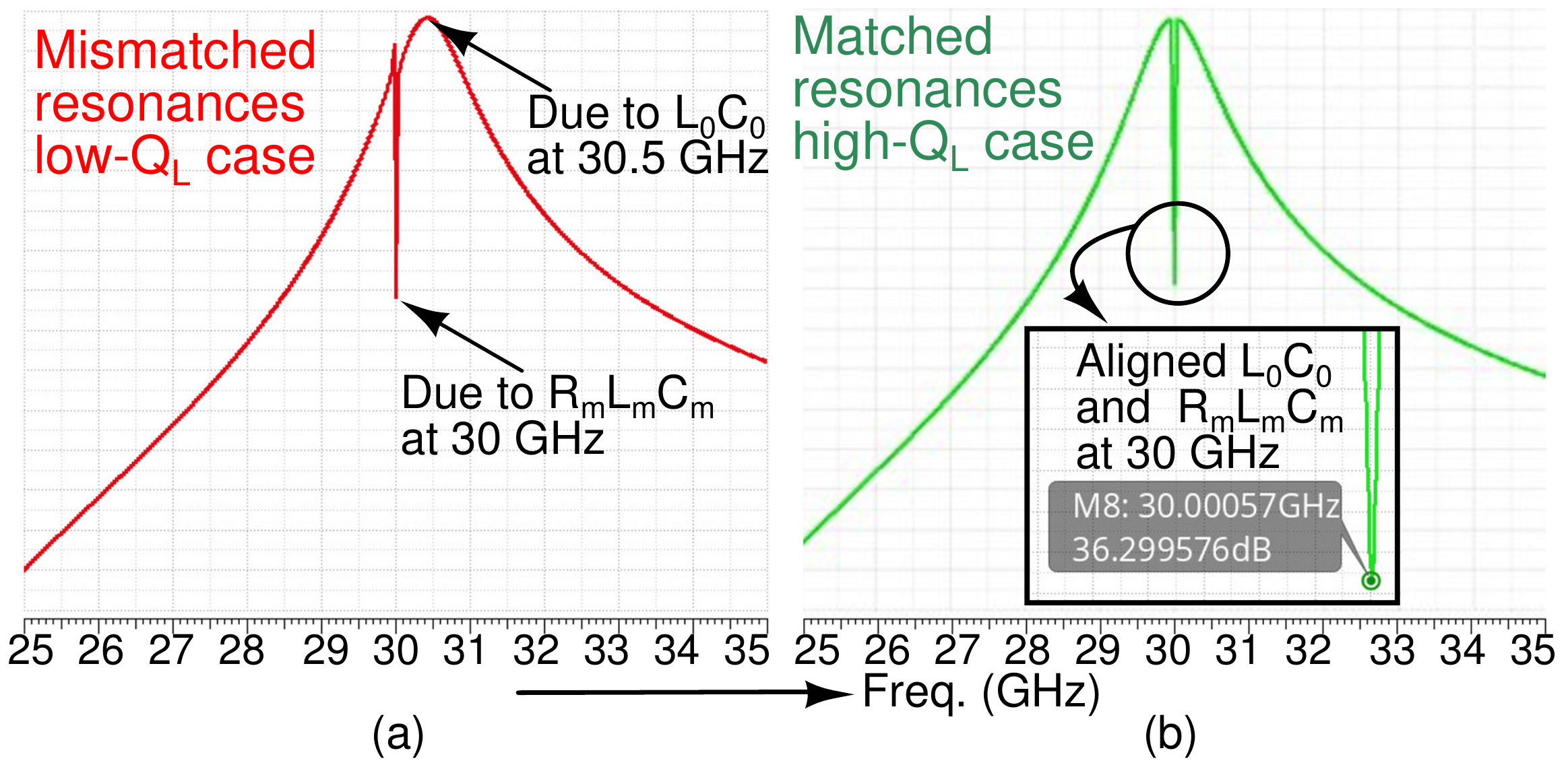}
\caption{ Impedance magnitude plot w.r.t frequency for cases with (a) mismatched and (b) matched resonances of $L_mC_m$ and $L_0C_0$ branches}
\label{fig_reso_mismatch}
\end{figure}
As discussed in previous section, $Q_L\approx Q_{RFT}$ (for $Q_{RFT}>2000$). 
However, due to the parallel combination of impedances, the shunt-inductor compensated resonator experiences a reduced effective resistance ($R_{RES}$) at resonance given by Eq. (\ref{eq_r_reso}).  
\begin{equation}
\label{eq_r_reso}
\small
R_{RES} = R_m || (Q_{L_0}^2\times R_{L_0})
\end{equation}
For a fixed bias current ($I_{BIAS}$), reduced $R_{RES}$ decreases both the oscillation amplitude ($V_{OSC} = I_{BIAS}\times R_{RES}$) and the carrier power $(P_{c} \propto V_{OSC}^2)$ in the current limited regime. 
From (\ref{eq_leeson}), $\mathcal{L}\{\Delta f\} \propto \frac{1}{Q_L^2P_c}$, which implies that high $Q_L$ will improve the phase noise whereas, reduced $P_c$ will degrade the phase noise. 
Therefore, overall phase noise can improve if high value of $Q_L$ dominates the reduced $P_c$. 
For example, as compared to a parallel LC tank with Q-factor of 20, with the shunt inductor compensated resonator with $Q_L = 2000$, if $P_c$ reduces by 20 dB due to reduced $R_{RES}$, 
there will still be about $20$ dB improvement in the phase noise. 
However, in order to get the advantage of high $Q_L$ in phase noise improvement, the resonant frequency of $L_0C_0$ branch should match closely to $f_s$.  
As shown in Fig. \ref{fig_reso_mismatch}(a), if there is significant mismatch, then due to the higher loop gain, oscillator will work at frequency defined by $L_0C_0$ tank with lower $Q_{L_0}$ and hence will have a poor phase noise. Whereas, if the two resonances are aligned (Fig.  \ref{fig_reso_mismatch}(b)) at $f_s$, $Q_L$ will be equal to $Q_{RFT}$ and very low phase noise can be obtained. 
\subsection{Quantitative Analysis}
\begin{figure}
\centering
\includegraphics[scale=0.46]{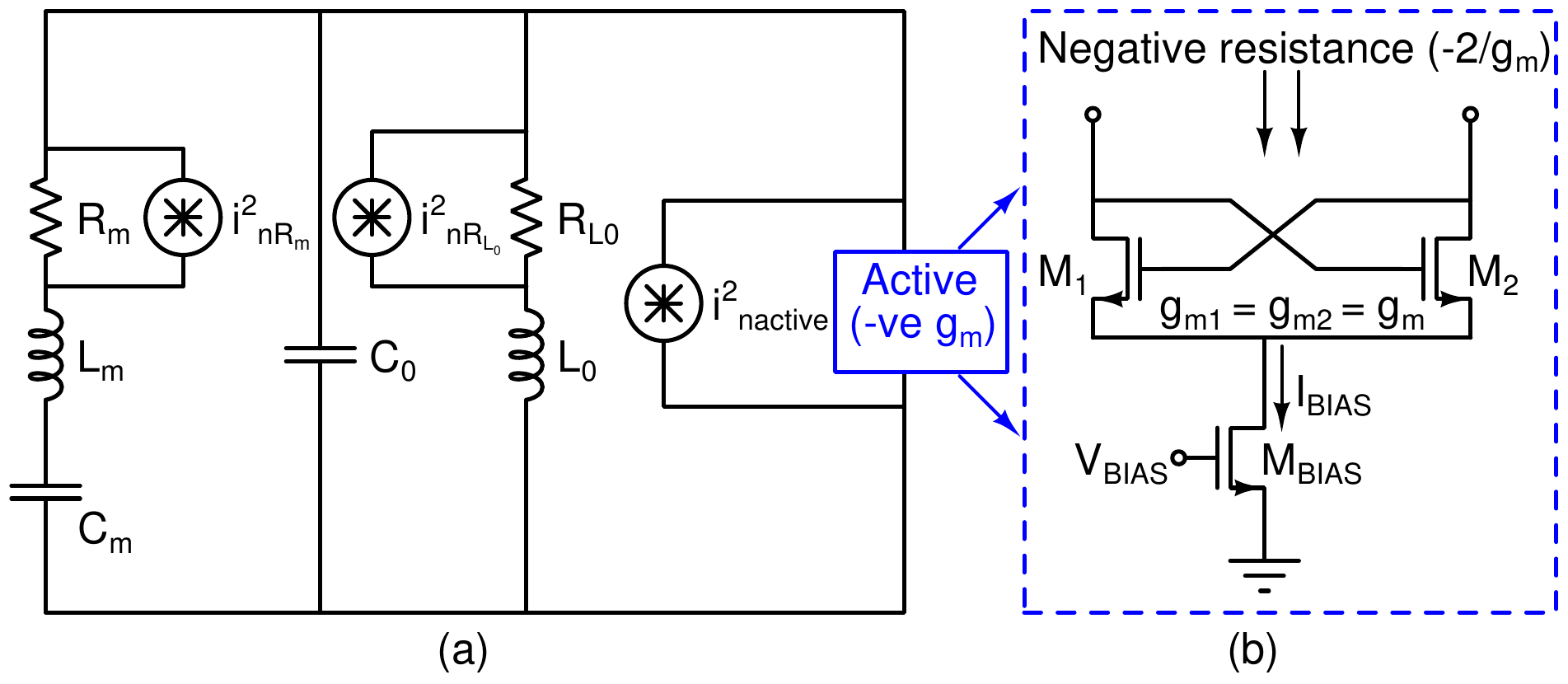}
\caption{(a) Depiction of different noise sources in the circuit (b) active circuit providing negative resistance for oscillation}
\label{fig_pn_sources}
\end{figure}

As shown in Fig. \ref{fig_pn_sources}(a), the oscillator 
has 3 major noise sources- 1) $R_m$, 2) $R_{L_0}$ and 3) active part of the circuit providing negative conductance for sustained oscillations (Fig. \ref{fig_pn_sources}(b)). 
The fundamental minimum noise factor ($F_{min}$) of the oscillator is given by Eq. (\ref{eq_f_min}).
\begin{equation}
\label{eq_f_min}
\small
F_{min} = 1 + F_{R_{L_0}} + F_{ACTIVE}
\end{equation}
where, $F_{R_{L_0}}$ and $F_{ACTIVE}$ are the noise factors due to $R_{L_0}$ and non-linear active circuit, respectively. 
$F_{R_{L_0}}$ and $F_{ACTIVE}$ %
can be represented by expressions (\ref{eq_f_active}a) and (\ref{eq_f_active}b) extending the definitions presented in \cite{pn_process_abidi_2000} and \cite{hegazi_jssc_2001} for an LC oscillator.  
\begin{subequations}
\label{eq_f_active}
\begin{align}
\small
&F_{R_{L0}} = \frac{R_{L_0}}{R_m}\\
\small
&F_{ACTIVE} = \gamma\frac {R_{RES}}{R_m} + \gamma \frac{4}{9}g_{mbias} R_m\bigg(\frac{R_{RES}}{R_m}\bigg)^2
\end{align}
\end{subequations}
where, $g_{mbias}$ is the transconductance of the tail current source and $\gamma$ is the channel noise coefficient of FET. By combining expressions (\ref{eq_f_min}) and (\ref{eq_f_active}) and defining $\beta = \frac{R_{RES}}{R_m}$ we get Eq. (\ref{eq_f_min_final}) shown below. 
\begin{equation}
\label{eq_f_min_final}
\small
F_{min} = 1 +\frac{R_{L_0}}{R_m} + \gamma \beta + \gamma \frac{4}{9}g_{mbias}R_m\beta ^2
\end{equation}
Eq. (\ref{eq_f_min_final}) gives the fundamental minimum noise factor of inductively compensated MEMS resonator based mmWave oscillator. 
Eq. (\ref{eq_f_min_final}) along with Eq. (\ref{eq_leeson}) give important insights about the design choices for oscillator. In order to reduce the $F_{min}$, lower values of $R_{L_0}$ and  $\beta$ are desirable. However, it is also desirable to have highest possible $Q_L$ and hence to have maximum $Q_{L_0}$ (=$\frac{\omega L_0}{R_{L_0}}$) to have least effect on $Q_{RFT}$ as discussed in section \ref{c0_compensation}. From Eq. (\ref{eq_r_reso}), $\beta$ can be reduced by choosing lower value of $R_{L_0}$, however with limited $Q_{L_0}$, it is only possible if $L_0$ is reduced. 
This leads us to a general suggestion to use a lower $L_0$ with lower $R_{L_0}$ with highest possible $Q_{L_0}$. 
This suggestion is counter intuitive as in LC oscillators we try to maximize ${L_0}$ such that higher oscillation amplitude can be achieved and hence reduced phase noise. 
\begin{figure}%
\centering
\includegraphics[scale=0.245]{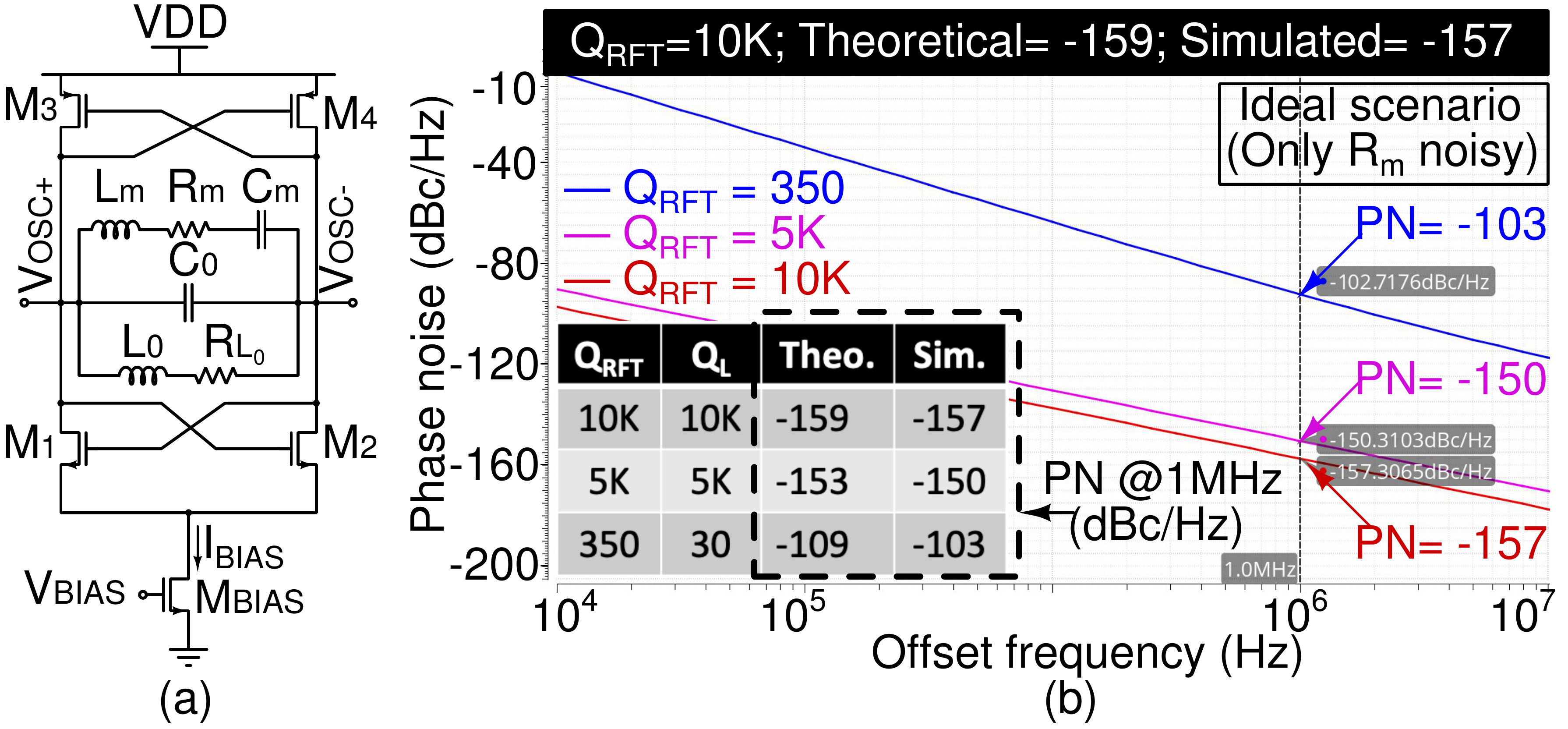}
\caption{(a) Oscillator schematic, (b) simulation results showing phase noise (dBc/Hz) at 1 MHz offset near 30 GHz for different values of $Q_{RFT}$, while considering noise due to $R_m$ only}
\label{fig_pn_ideal}
\end{figure}

As compared to LC oscillator, F increases for the proposed shunt-inductor compensated MEMS based oscillator by a factor of $\frac{R_{L_0}}{R_m}$, however, $\frac{F}{Q_L^2}$ reduces considerably due to very high $Q_L$ of the resonator as discussed in section \ref{c0_compensation}. 
Theoretical minimum phase noise for 100\% ideal oscillator ($F_{min} = 1$) can be estimated by considering noise due to $R_m$ only. 
For $R_m$ = 332 $\Omega$, $Q_{RFT}$ = 10K, $V_{OSC}$ = 300  mV, $L_0$ = 250 pH, $R_{L_0}$ = 4.8 $\Omega$, $Q_{L_0}$ = 10, $\beta$ = 0.6, the calculated phase noise is about -159 dBc/Hz at 1 MHz offset for $f_0$ = 30 GHz. 
For validating the theoretical values, a cross-coupled oscillator is designed in 14 nm GF technology 
(Fig. \ref{fig_pn_ideal}(a)). %
From simulations, while keeping only the noise contribution of $R_m$ ON and with the same component values used in theoretical analysis, the phase noise is about -157 dBc/Hz for $Q_{RFT}$ = 10K, which matches closely with the theoretical value.  

FoM of an oscillator can be defined as follows:
\begin{equation}
\label{eq_fom}
\small
FoM = \frac{\big(\frac{f_0}{\Delta f}\big)^2}{\mathcal{L}\{\Delta f\} P_{DC}[mW]}
\end{equation}
From equations (\ref{eq_leeson}) and (\ref{eq_fom}) FoM for the shunt-inductor compensated MEMS resonator based oscillator can be defined as follows:
\begin{equation}
\label{eq_fom_mems_osc}
\small
FoM = \frac{Q_L^2}{KTF}\times \frac{V_{OSC}^2}{R_mP_{DC}} \times 10^{-3}
\end{equation}
Defining the power dissipated at output as $P_{OUT} = \frac{V_{OSC}^2}{2R_{RES}}$ and oscillator efficiency as $\eta = \frac{P_{OUT}}{P_{DC}}$, Eq. (\ref{eq_fom_mems_osc}) reduces to $FoM = 2\beta \eta\frac{Q_L^2}{KTF}\times 10^{-3}$. 
Theoretical maximum FoM of 100\% efficient oscillator ($\eta$=1), with noise-less $R_{L_0}$ and noise-less negative conductance (F=1), can be given by Eq. (\ref{eq_fom_mems_osc_db}).
\begin{equation}
\label{eq_fom_mems_osc_db}
\small
FoM_{max} = 176.8 + 20\,logQ_L + 10\,log \beta\ (dBc/Hz)
\end{equation}
The value of $FoM_{max}$ is about 250 dBc/Hz for the values taken in the previous example to calculate theoretical minimum phase noise, which is about 50 dB more than that for the conventional LC oscillators \cite{hegazi_jssc_2001}. 
\subsection{General Design Technique for mmWave Oscillator with on-chip MEMS Resonator}
\label{design_tech}
\begin{itemize}
\item Plot magnitude and phase of the impedance from the electrical model of the on-chip MEMS resonator. 
\item Identify the value of shunt inductor ($L_0$) to compensate $C_0$ taking into account the routing parasitics and buffer load at the oscillator output node. 
\item In order to minimize $R_{L_0}$ and $\beta$, use minimum value of $L_0$ with 
highest possible $Q_{L_0}$ for $Q_{RFT} \geq 2000$. 
Use additional MIM cap ($C_{fix}$) parallel to $L_{0}$.
\item Use capacitor bank with very small unit size ($\approx 1 fF$) to add frequency tunability to exactly resonate out $C_0$ with $L_0$ at $f_s$ for the best phase noise performance. 
\item Use minimum gate length transistors for mmWave speed and reduced area to implement the active negative conductance required for the oscillator. 
\item For initial sizing of the active part, find minimum transconductance using $g_m = \frac{2}{R_{RES}}$, current using $I_{BIAS} = \frac{V_{OSC}}{R_{RES}}$ and then calculate $W/L = (g_m)^2/2(I_{BIAS}\mu _{n,p}C_{ox})$. Simulate and modify the design as needed. 
\end{itemize}

\section{Implementation and Simulation Results}
\label{impl_result}
\begin{figure}%
\centering
\includegraphics[scale=0.255]{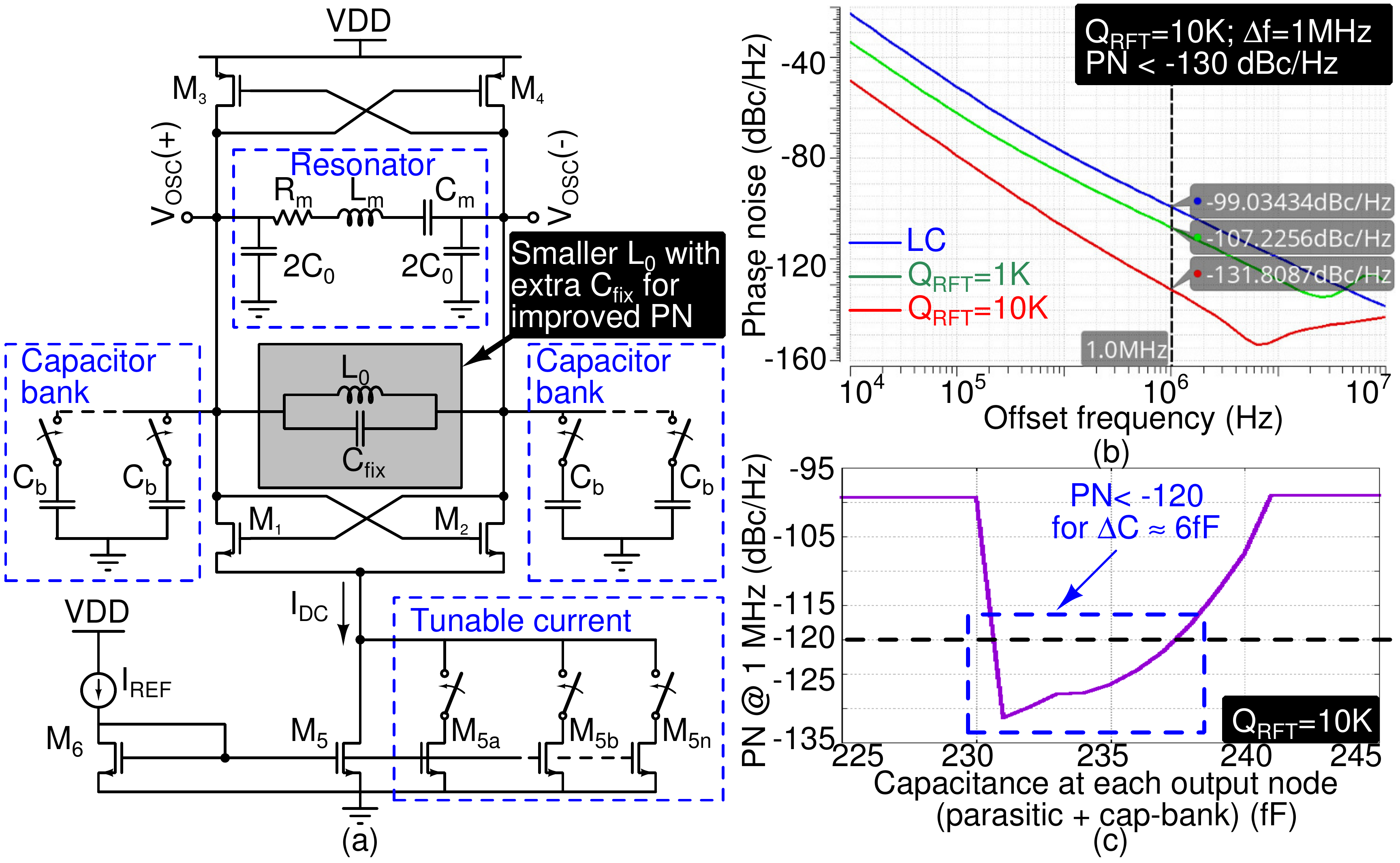}
\caption{(a) Schematic of the proposed oscillator, (b) post layout simulation results comparing phase noise of LC and the proposed oscillator, (c) phase noise vs capacitance plot depicting sensitivity to resonance matching}
\label{fig_passive_osc}
\vspace{-4mm}
\end{figure}

Fig. \ref{fig_passive_osc}(a) shows the schematic of the proposed shunt-inductor compensated oscillator topology, which has been implemented in 14 nm Global Foundry (GF) 12LPP technology following the guidelines  proposed in section \ref{design_tech}. 
A MOS capacitor bank with unit size of $C_b$ ($\approx 1fF$) with a fixed MIM capacitor ($C_{fix}$) of 10 fF is added in parallel with $L_0$ to achieve a fine tuning of the tank for exact alignment of $L_0C_0$ and the motional branch ($R_mL_mC_m$) to achieve the highest loaded $Q_L$. 
An inductor of 250 pH ($Q \approx 8$) has been used from the PDK to realize $L_0$.
Post-layout simulation results (Fig. \ref{fig_passive_osc}(b)) show that phase noise of the proposed oscillator is  $-132$ dBc/Hz at an offset of 1 MHz for the carrier frequency of 30 GHz for $Q_{RFT}$ = 10,000. 
Fig. \ref{fig_passive_osc}(c) shows the sensitivity of phase noise with respect to the non-alignment of the two resonances. Capacitance at output is changed to shift the $L_0C_0$ resonance frequency. As shown in the figure, for about 6 fF variation in output capacitance, effect of high $Q_L$ is present and phase noise is $<-120$ dBc/Hz at 1 MHZ offset. 
The proposed circuit consumes about 2 mW power from 0.8 V supply and have a simulated FoM of 217 dBc/Hz. 
As shown in table \ref{tab_results}, the proposed MEMS oscillator achieves $>$25 dB improvement in the FoM, while consuming extremely low power as compared to the other recently reported LC oscillators. 
\begin{table}[t]
\begin{center}
  \scriptsize
 \caption{Performance summary and comparison}
  \vspace{-6mm}
  \label{tab_results}
      \resizebox{1\linewidth}{!}{
  \begin{tabular}{|>{\centering}p{1.23cm}|>{\centering}p{1.1cm}|>{\centering}p{1.1cm}|>{\centering}p{1.1cm}|>{\centering}p{1.03cm}|>{\centering}p{1.09cm}|}
        \hline
         \textbf{Parameters} &\textbf{\cite{lco_martin_isscc_2019}} & \textbf{\cite{lco_nagendra_isscc_2019}} & \textbf{\cite{lco_shu_rfic_2018}} & \textbf{\cite{lco_fabio_isscc_2018}}  & \textbf{This work}\tabularnewline
        \hline
         Measured/ Simulated & Measured & Measured & Measured & Measured & Simulated \tabularnewline
        \hline
         Technology & $65$ nm CMOS & $65$ nm CMOS & $28$ nm CMOS& 135 nm BiCMOS & $14$ nm CMOS\tabularnewline
         \hline        
        Frequency & 29.92 GHz & 28 GHz &25.56 GHz  & 15 GHz& 30 GHz\tabularnewline
        \hline
         Resonator & LC & LC & LC   & LC & MEMS  \tabularnewline
        \hline	
        Oscillator Phase Noise & -112.3 dBc/Hz & $> -110$ dBc/Hz \dag & -102.4 dBc/Hz  & -124 dBc/Hz & -132 dBc/Hz \tabularnewline
        (offset) & (1 MHz) & (1 MHz) & (1 MHz)  & (1 MHz) & (1 MHz) \tabularnewline
		\hline
         Power & 4 mW & $17.5$ mW & 5.5 mW   & 72 mW & 2 mW  \tabularnewline
        \hline		
        FoM (dBc/Hz) & 189.8 & 187 & 183.3 &189 & 217\tabularnewline
        \hline          
         Supply & $0.48$ V & $0.65$ V &  $0.9$ V & $3$ V & $0.8$ V\tabularnewline
        \hline         
  \end{tabular}
  }
\begin{tablenotes}
\item \dag Taken from the graph in the paper., where $-117$ dBc/Hz is mentioned at 3 MHz
 \end{tablenotes}
\end{center}
\end{table}

\section*{Conclusions}
In this work, insights to achieve fundamental limits of phase noise in a mmWave oscillator circuit with high-Q on-chip MEMS resonator with extremely small footprint have been presented with detailed analysis. 
Following the analysis, general oscillator design technique has been proposed 
while providing solutions to the problems due to the high static capacitance of high-Q on-chip MEMS resonator. 
To validate the proposed analysis and low phase noise design method, a 30 GHz oscillator has been designed in 14 nm GF 12LPP technology. Post layout simulations show that 30 GHz oscillator exhibits a phase noise of -132 dBc/Hz at 1 MHz offset, while consuming 2 mW power from 0.8 V supply. The simulated FoM of the oscillator is 217 dBc/Hz, 
which is $>$25 dB better than the existing mmWave LC oscillators near 30 GHz.  
\addtolength{\textheight}{-12cm} %

\section*{Acknowledgement}
Authors acknowledge DARPA MIDAS program for funding this research work.
\bibliography{ref_mmwave_osc}
\end{document}